\documentclass[]{article}  
\usepackage{graphicx}              

\begin{document}

\title{THE HIGGS DISCOVERY POTENTIAL OF ATLAS}   
\author{CHRISTOPHER COLLINS-TOOTH\\
\textit{for the ATLAS collaboration}\\
\\
\textit{Department of Physics and Astronomy, University of Glasgow,}\\
\textit{Glasgow G12 8QQ, Scotland.}
}                       

\date{October 2, 2007}                           
\maketitle

Higgs boson production and decay at the LHC is described, together with related 
ATLAS search channels, in order to provide an overview of the ATLAS Higgs discovery potential.
\newline
\newline

\section{Introduction}
The ATLAS detector \cite{key:ATLASTDR1} is one of two general purpose detectors 
under construction at the Large Hadron Collider (LHC), CERN, Switzerland. 
The LHC will begin to provide ATLAS with colliding beams, each of $7$~TeV protons, 
in the Summer of 2008. It is expected that a low-luminosity running phase will commence 
in 2009, with instantaneous luminosity ($L$) of $1-2$~cm$^{-2}$s$^{-1}$. The aim of this phase will
be to gather approximately $30$~fb$^{-1}$ of integrated luminosity ($\int L dt$) by 2011. After this,
$L$ will be increased to approximately $10^{34}$~cm$^{-2}$s$^{-1}$, with 
the aim of gathering $\int L dt~ \sim300$~fb$^{-1}$ by 2014/2015.
The ATLAS trigger will reduce the frequency of events saved for 
offline analysis to approximately $\sim200$~Hz. 

Some important design features of ATLAS are its high hermeticity (vital in measurements of 
missing transverse energy), powerful tracking in the Inner Detector (useful for example, 
in b-tagging measurements), excellent Electromagnetic Calorimeter energy and angular resolution 
(e.g. for measurement of $\gamma,~e^{\pm}$ energies, and separation of 
$\gamma$/jet \& $\gamma$/$\pi^0$). ATLAS is also designed to have excellent muon detection 
efficiency and momentum measurement, using the Inner Detector and the Muon Spectrometer. 
The nominal energy scales are $e/\gamma \sim 0.1\%, \mu \sim 0.1\%, jets \sim 1\%$.

\section{Standard Model Higgs production and decay at the LHC}
\begin{figure}[htbp]
\vfill
\begin{center}
$\begin{array}{cc}
\includegraphics[width=6.2cm]{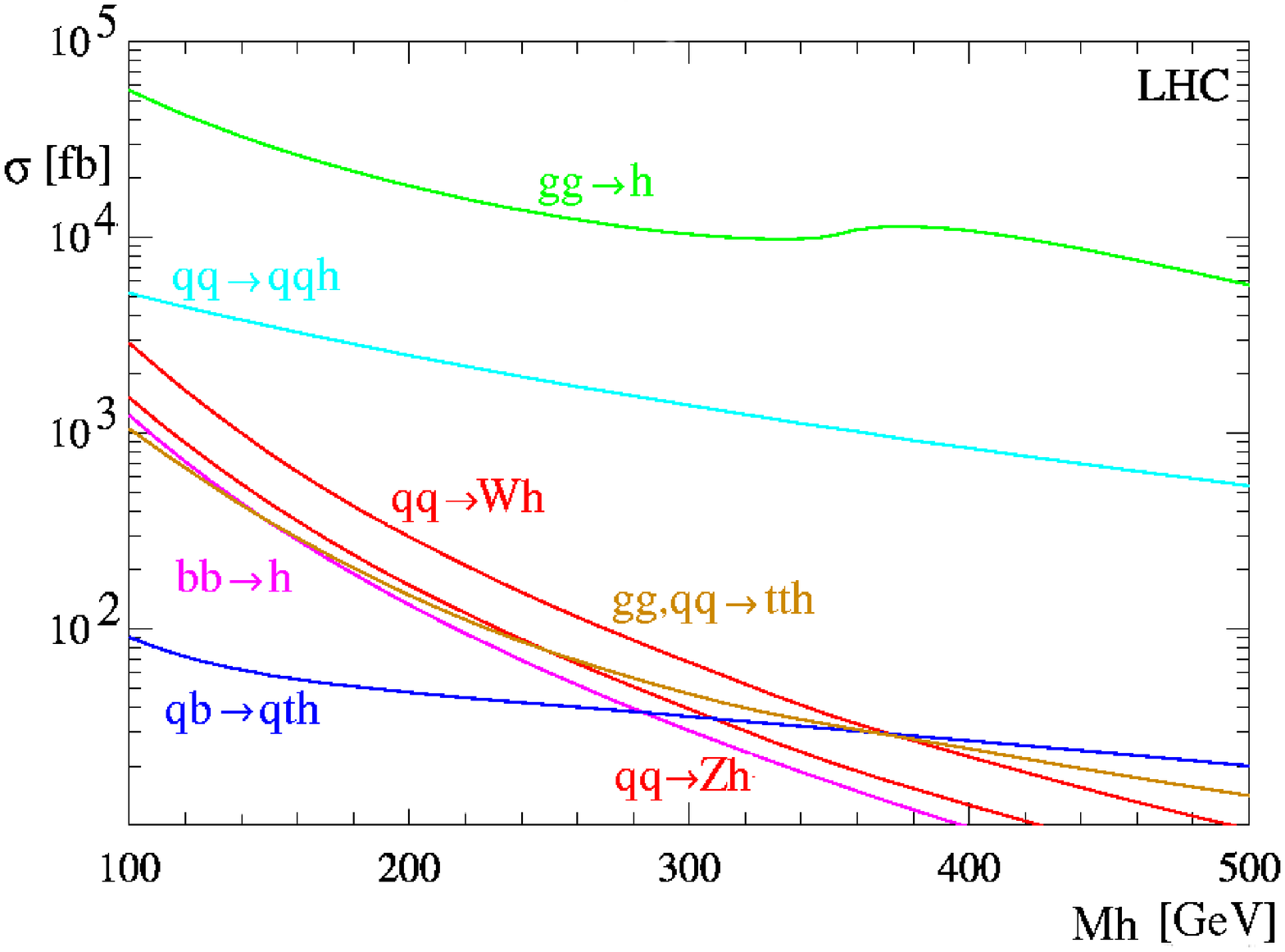}
\includegraphics[width=6.3cm]{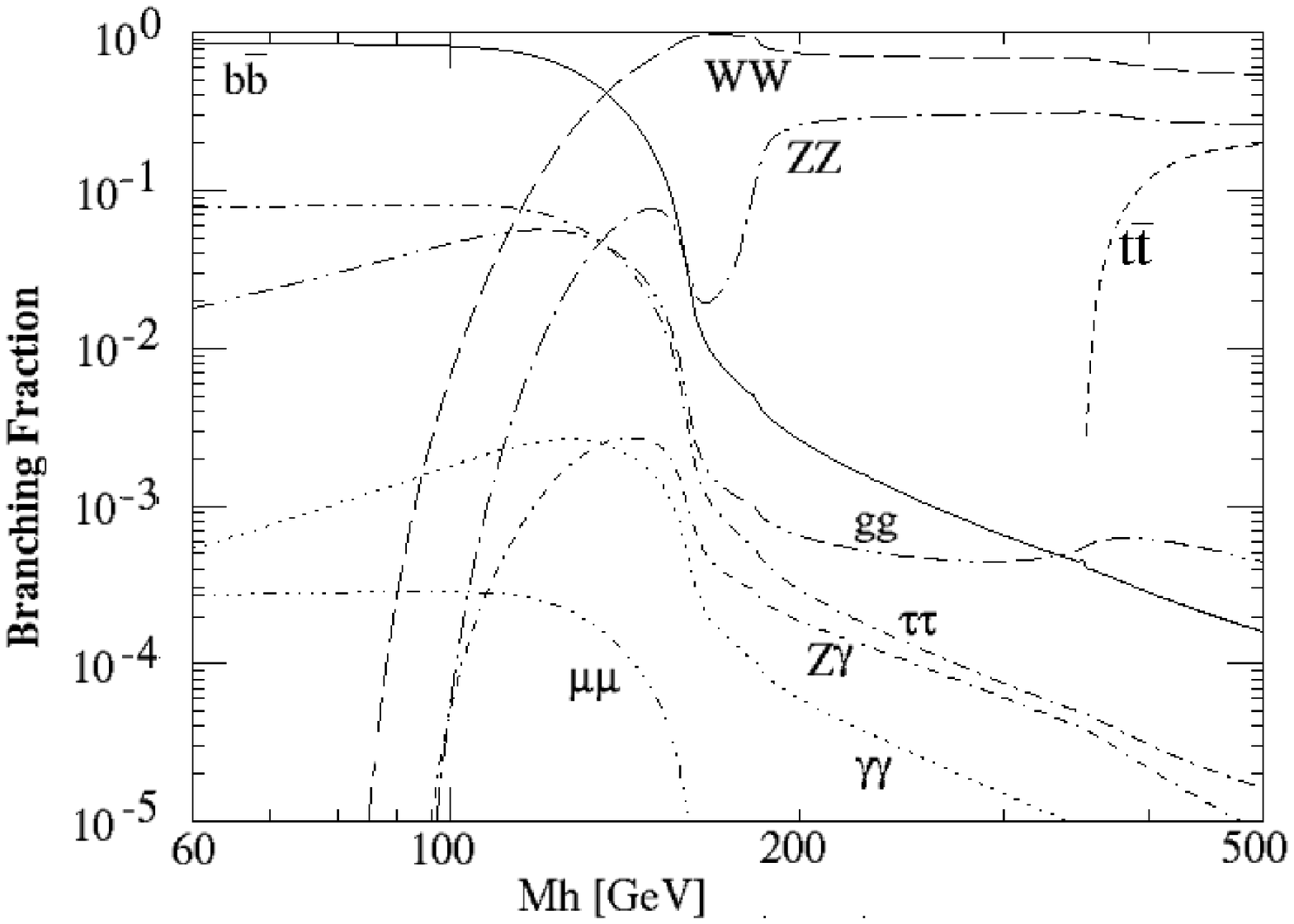}
\end{array}$
\end{center}
\caption{
SM Higgs production (left) \cite{key:TeV4LHCHiggsSMProd} and decay (right) \cite{key:TeV4LHCHiggsSMDecay} at the LHC. 
}
\label{key:fig-SMHProductionDecay}
\vfill
\end{figure}

Figure \ref{key:fig-SMHProductionDecay} shows that by far the most important production mechanism over the 
entire mass search region for a Standard Model (SM) Higgs boson is gluon-gluon fusion. This production
mechanism is followed in importance by Vector Boson Fusion (VBF) at the $10-20\%$ level, and
Z/W-associated \& associated-top production at a still lower cross-section. 

The branching ratios for SM Higgs decay are also shown in Figure
\ref{key:fig-SMHProductionDecay}. In the Higgs mass region $M_h <125$~GeV, the primary decay mechanism is
$h\rightarrow b\bar{b}$, At higher masses, decays to (initially off-shell) gauge bosons become dominant.
This occurs first for $W$-bosons, and this decay stays dominant over the decay to $Z$-bosons, though the latter is
preferred at higher masses ($M_h > 2M_Z$), because it has a very clean experimental signature.

\section{ATLAS Higgs sensitivity and searches}
\subsection{ATLAS SM Higgs sensitivity}
ATLAS is optimised to cover a large spectrum of possible Higgs particles and their signatures.
The ATLAS SM Higgs significances for the individual channels as a function of $M_h$ are shown in Figure 
\ref{fig:AtlasSignificances}, and also shown is the overall significance when multiple channels 
are considered. ATLAS should make a light Higgs discovery ($M_h < 180$~GeV) with $\int Ldt=30$~fb$^{-1}$.
Discovery in the region $M_h<120$~GeV is clearly the most challenging for ATLAS, and 
will probably require several channels to be combined. Given a favourable Higgs mass, ATLAS could make a discovery
with much less luminosity. By combining VBF, $h\rightarrow\gamma\gamma$, $t\bar{t}h\rightarrow b\bar{b}$ \&
$ZZ^*$, discovery of a Higgs with $M_h > 120$~GeV could be made with just $10$~fb$^{-1}$ of 
luminosity \cite{key:ATLASSignificances}. Above $M_h > 2M_Z$, two on-shell $Z$-bosons can be created, opening
up the `gold plated' channel $h \rightarrow ZZ \rightarrow 4l^{\pm}$.

\subsection{$h \rightarrow \gamma\gamma$}
\begin{figure}
  \begin{minipage}[b]{0.45\linewidth} 
    \centering
    \includegraphics[width=5.5cm]{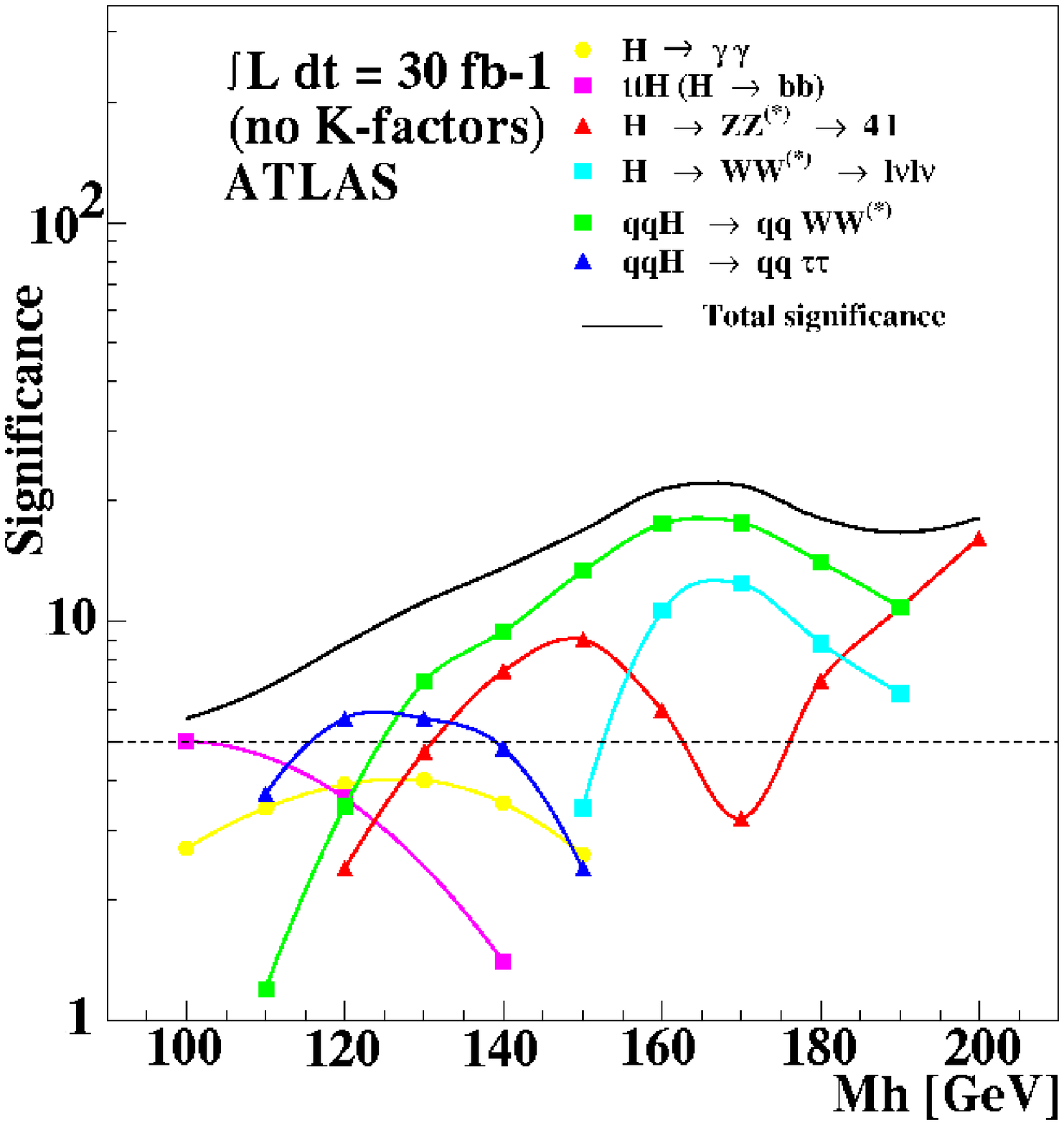}
    \caption{ATLAS sensitivity to the discovery of a SM Higgs boson for $\int Ldt=30$~fb$^{-1}$ \cite{key:ATLASSignificances}.
      No K-factors have been applied.}
    \label{fig:AtlasSignificances}
  \end{minipage}
  \hspace{0.5cm} 
  \begin{minipage}[b]{0.45\linewidth}
    \centering
    \includegraphics[width=5.7cm]{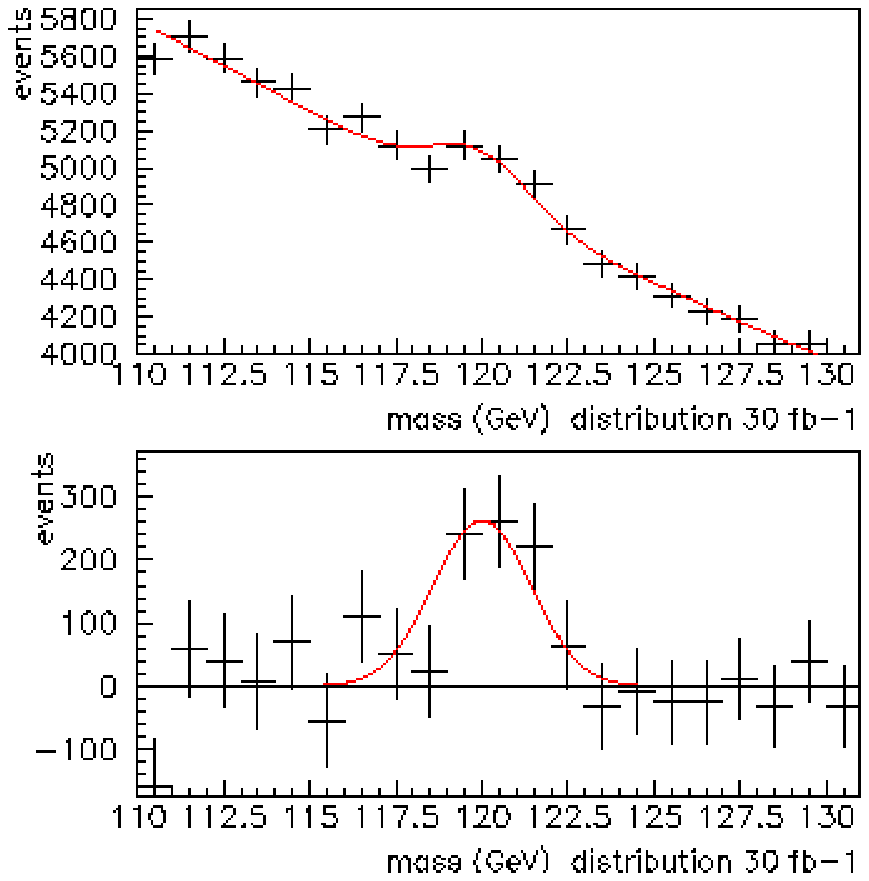}
    \caption{Reconstructed $M_h$ for an input $M_h=120$~GeV before (top) and after (bottom) background subtraction
      ($\int Ldt=30$~fb$^{-1}$) \cite{key:ATLASHgg}.}
    \label{fig:ATLASHgg}
  \end{minipage}
\end{figure}

This channel features a narrow mass peak over a smooth background, but as can be seen from 
Figure \ref{fig:AtlasSignificances}, it is only useful at low $M_h$ ($<140$~GeV). This is a benchmark channel for 
ATLAS detector performance, relying upon Electromagnetic Calorimeter resolution, but also accurate 
primary vertex determination. The irreducible background to this channel is the $\gamma\gamma$
continuum. Powerful rejection is needed to reduce jet-jet \& $\gamma$-jet backgrounds,
where the jet could be misidentified as a photon. When the analysis is performed using NLO production 
cross-sections \cite{key:ATLASHgg}, the ATLAS significance varies from 5.4 ($M_h=140$~ GeV), 
through to 6.2 ($M_h=130$~GeV) and 6.1 ($M_h=120$~GeV) for $30$~fb$^{-1}$ of integrated luminosity.
Figure \ref{fig:ATLASHgg} shows a mass distribution for an assumed $M_h=120$~GeV. The analysis 
makes use of photon conversion recovery, where for the approximately 40\% of signal events where at least 
one photon converts, an 80\% recovery rate is assumed. 

\subsection{$h \rightarrow ZZ^* \rightarrow 4l^{\pm}$}
This channel has a clean signature, but low statistics ($3-11$~fb for $M_h=130-200$~GeV) \cite{key:ATLASZZ4l}.
It is another benchmark channel for detector performance, with the appearance of four high transverse momentum ($P_T$)
charged leptons testing the ATLAS momentum and energy resolution (via the Inner Detector, Muon Spectrometer
and Electromagnetic Calorimeter).
Mass peaks can be produced in this channel e.g. for $4e,~4\mu,~2e2\mu$ final states, which differ by resolution,
but are typically in the range $1.5-2$~GeV. The channel also provides a possibility of determining the spin
and CP-eigenvalue of a SM Higgs \cite{key:ATLASZZ4lSpin}. The main backgrounds to the channel
are the irreducible $ZZ^* \rightarrow 4l^{\pm}$ continuum, and the (reducible)
$t\bar{t}\rightarrow 4l+X$ \& $Zb \bar{b} \rightarrow 4l +X$. These backgrounds
contain non-isolated leptons with high impact parameters. Once selection is performed, the $ZZ$-continuum
background is dominant. The shape and normalisation of these backgrounds will be obtained experimentally
from data in order to minimise PDF and luminosity uncertainties. The ATLAS signal significance with 
no K-factors applied (for $\int Ldt = 30$~fb$^{-1}$) rises above $5\sigma$ at $M_h \approx 130$~GeV, 
and increases to approximately $9\sigma$ at $M_h \approx 150$~GeV. The significance then drops 
below $5\sigma$ at $M_h \approx 160$~GeV \cite{key:ATLASZZ4l}.

\subsection{Vector Boson Fusion $h\rightarrow WW^{(*)}$}
In addition to being the second most important SM Higgs production mechanism after gluon-gluon fusion, 
VBF has a number of topological features which enable ATLAS to efficiently 
reject background. Due to this, the VBF channel $h \rightarrow WW^{(*)}$ in particular demonstrates large discovery 
potential over the region $125 < M_h < 190$~GeV. VBF produces Higgs decay products which lie between
two forward jets. These jets, with Higgs decay products located between them in pseudorapidity ($\eta$), 
coupled with an absence of jets in the central region (due to the absence of quark-quark colour exchange) 
provide the experimental signature of VBF. The two forward `tag jets' typically have higher invariant 
dijet mass ($M_{jj}$) than the tag jets identified in QCD background processes, and so a cut on $M_{jj}$ 
is also applied by ATLAS to improve background rejection. 

\begin{figure}[htbp]
  \begin{minipage}[b]{0.45\linewidth} 
    \centering
    \includegraphics[width=5.5cm]{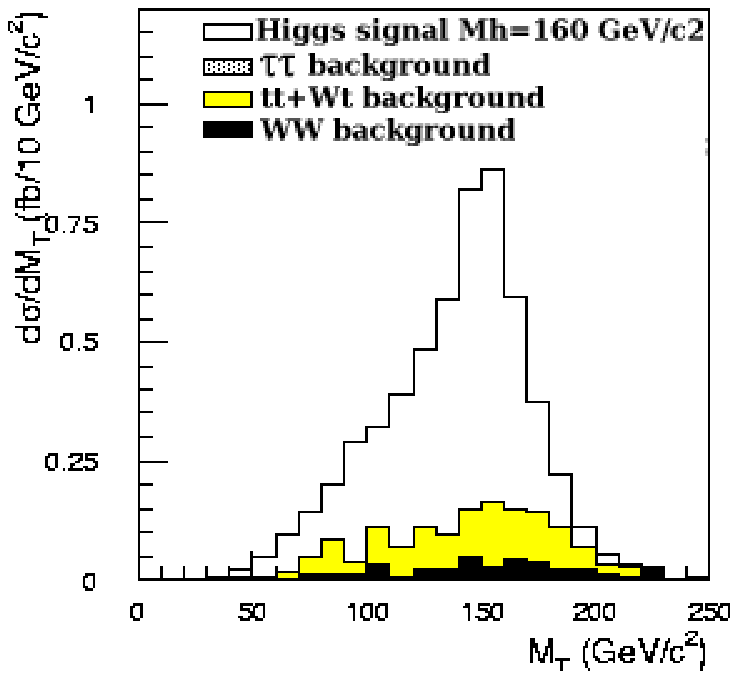}
    \caption{Distribution of the transverse mass $M_T$ for a SM Higgs boson mass of $160$~GeV
      in the $h\rightarrow WW^{*} \rightarrow l\nu l\nu~ (e\mu)$ channel after all selection cuts are applied
      \cite{key:ATLASSignificances}.}
    \label{key:fig-MtH160WWlnln}
  \end{minipage} 
  \hspace{0.5cm} 
  \begin{minipage}[b]{0.45\linewidth}
    \centering 
    \includegraphics[width=6.4cm]{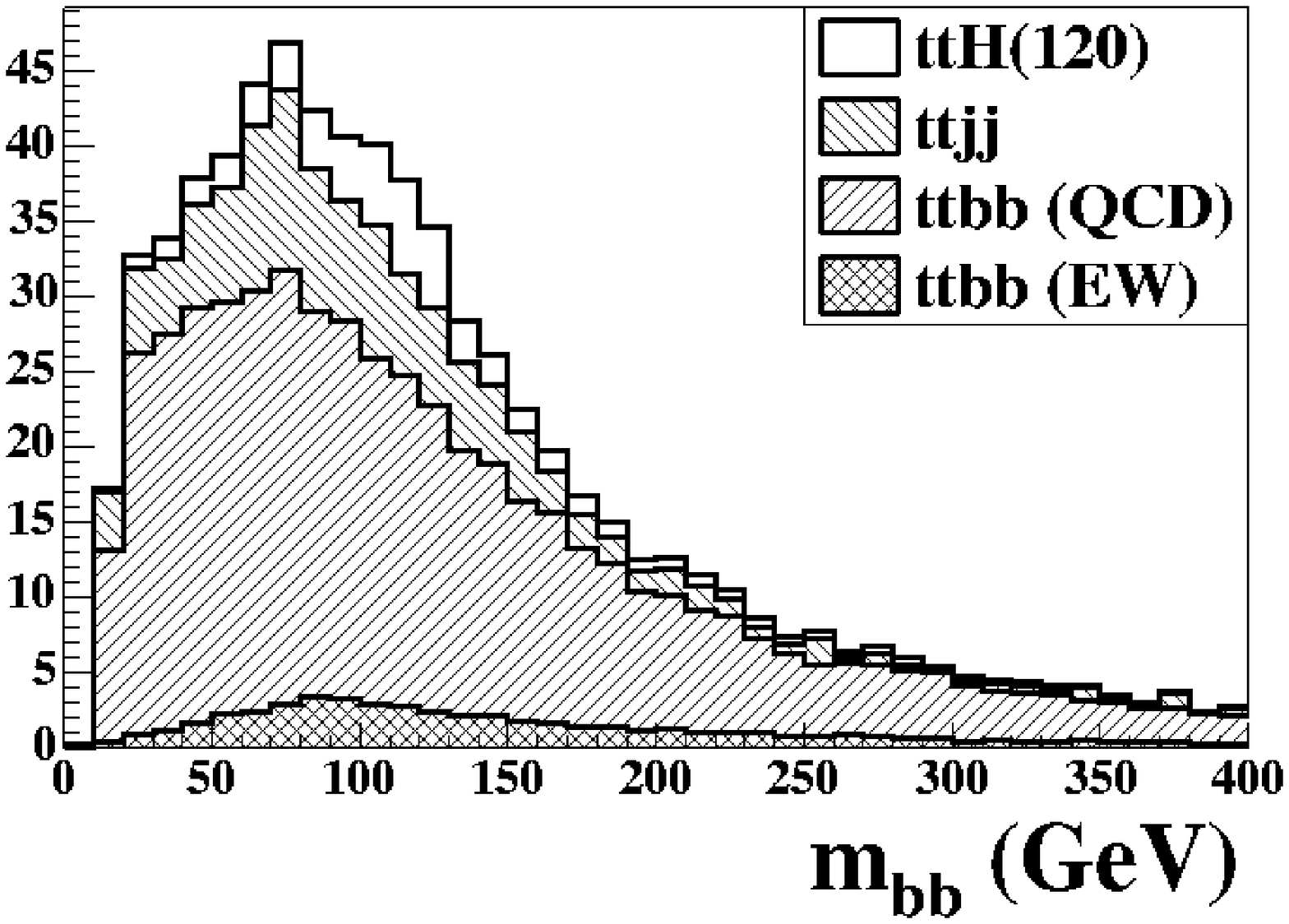}
    \caption{Mass distribution for the $t\bar{t}h~(h \rightarrow b\bar{b})$ semi-leptonic final state
      with relevant backgrounds for an assumed $M_h=120$~GeV and $\int Ldt=30$~fb$^{-1}$ \cite{key:ATLAStthhbb}.}
    \label{key:fig-ATLAStthhbb}
 \end{minipage}
\end{figure}

In the $h\rightarrow WW^{*}\rightarrow l\nu l\nu$ channel, the leptons are spin-correlated to the Higgs resonance. 
For a SM Higgs boson, the charged leptons will emerge in the same direction as each
other. This is yet another topological feature which improves background rejection. 
Due to the production of two neutrinos, only the transverse mass,
$M_T=\sqrt{}((E_T^{ll}+E_T^{\nu \nu})^2~-~(p_T^{ll}+p_T^{miss})^2)$ can be reconstructed.
This distribution for signal having $M_h=160$ in the $e\mu$ final state is shown in Figure 
\ref{key:fig-MtH160WWlnln}, along with the relevant background distributions. 

With $\int Ldt = 10$~fb$^{-1}$, the ATLAS signal significance at $M_h = 160$~GeV 
for the $l\nu l\nu$ final state reaches approximately $8.1 \sigma$ for the $e\mu$ channel,
and approximately $7.4 \sigma$ in the $ee/\mu \mu$ channel.   
The $l\nu jj$ final state is also considered by ATLAS, and provides a significance of $4.6$ at $M_h = 160$~GeV
for $\int Ldt = 30$~fb$^{-1}$.
With $\int Ldt = 10$~fb$^{-1}$, the ATLAS signal significance for the combination of considered 
$l\nu l\nu$ \& $l\nu jj$ channels is above $5\sigma$ for $135 < M_h < 190$~GeV \cite{key:ATLASSignificances}.

\subsection{$t\bar{t}h~(h \rightarrow b\bar{b})$}
This channel is attractive primarily due to the large branching ratio of $H\rightarrow b\bar{b}$ 
at $M_h<130$~GeV (see Figure \ref{key:fig-SMHProductionDecay}). It is also relatively simple to trigger 
if one considers the semi-leptonic final state $t\bar{t}h \rightarrow l\nu jj b\bar{b}b\bar{b}$, 
which provides a high-$P_T$ isolated lepton and missing transverse energy, together with four $b$-jets.
The channel has a complex final state, and thus there is a large combinatoric background which 
reduces the mass-peak resolution (Figure \ref{key:fig-ATLAStthhbb}). 
Additionally, there are several physics backgrounds to consider.
The $t\bar{t}jj$ background requires ATLAS to optimise the light jet rejection fraction when b-tagging jets,
whilst $t\bar{t}b\bar{b}$ appears generally through QCD gluon radiation (giving four $b$-jets in the final state).
Kinematic information can typically be used to differentiate this background from signal, because the 
extra $b$-jets are not from a Higgs boson. 

The analysis selects events with a high jet multiplicity ($\geq 6$), and at least four $b$-tagged jets. 
The missing transverse energy is used to reconstruct the final state neutrino. The two $W$-bosons 
and top-quarks are required to be reconstructed (to remove background), and finally the remaining $b$-jets
are combined to reconstruct the Higgs boson. The ATLAS analysis \cite{key:ATLAStthhbb} has a significance of $~2.4$
for $M_h = 120$~GeV, which falls as $M_h$ increases (due to the rapid decrease in $H\rightarrow b\bar{b}$ 
branching ratio).

\subsection{The MSSM Higgs discovery potential of ATLAS}
\begin{figure}[htbp]
\vfill
\begin{center} 
    \includegraphics[width=5.3cm]{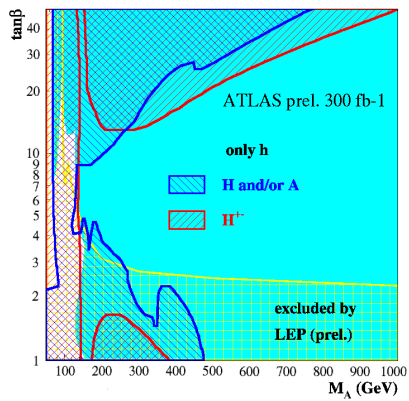}
\end{center}
\caption{ATLAS discovery potential for light and heavy neutral and charged Higgs bosons for
$\int Ldt=300$~fb$^{-1}$ in the $M_h$-max scenario \cite{key:ATLASMSSMmhmax}.}
\label{key:fig-mhmax300}
\vfill
\end{figure}

In the MSSM at Born level, the Higgs sector phenomenology is determined by two parameters, 
the vacuum expectation value ratio of the two Higgs doublets $\tan\beta$, and 
the Higgs mass $M_A$. The $M_h$-max scenario (where SUSY parameters give the largest value for $M_h$)
corresponds to the most conservative exclusion from LEP. The ATLAS discovery potential for this scenario
is shown in Figure \ref{key:fig-mhmax300}.
If $\tan\beta$ \& $M_A$ do in fact lie in the moderate $\tan\beta$ and large $M_A$ region of Figure 
\ref{key:fig-mhmax300}, then only one Higgs boson will be found. In this case, other observables such as the ratio 
$BR(h\rightarrow WW) / BR(h \rightarrow \tau \tau)$ will be used to infer the possible existence of a supersymmetric
Higgs sector.

\section{Conclusions}
The ATLAS experiment has discovery potential for a SM Higgs boson over the 
mass range defined by the LEP exclusion limit of $M_h>114.4$~GeV (95\% C.L.) 
\cite{key:MhLowerbound}, to $\sim 1$~TeV \cite{key:MhUpperbound}. 
A significant region of MSSM parameter space would yield discovery of multiple supersymmetric 
Higgs particles, should they exist.

\end{document}